# Measuring Bangladeshi Female Farmers' Values for Agriculture Mobile Applications Development

Rifat Ara Shams
Faculty of IT
Monash University
Melbourne, Australia
rifat.shams@monash.edu

Mojtaba Shahin
Faculty of IT
Monash University
Melbourne, Australia
mojtaba.shahin@monash.edu

Gillian Oliver
Faculty of IT
Monash University
Melbourne, Australia
gillian.oliver@monash.edu

Waqar Hussain
Faculty of IT
Monash University
Melbourne, Australia
Waqar.Hussain@monash.edu

Harsha Perera
Faculty of IT
Monash University
Melbourne, Australia
Harsha.Perera@monash.edu

Arif Nurwidyantoro
Faculty of IT
Monash University
Melbourne, Australia
Arif.Nurwidyantoro@monash.edu

Jon Whittle
CSIRO's Data61
Melbourne, Australia
Jon.Whittle@data61.csiro.au

**Abstract**

*The ubiquity of mobile applications (apps) in daily life raises the imperative that the apps should reflect users' values. However, users' values are not usually taken into account in app development. Thus there is significant potential for user dissatisfaction and negative socio-economic consequences. To be cognizant of values in apps, the first step is to find out what those values are, and that was the objective of this study conducted in Bangladesh. Our focus was on rural women, specifically female farmers. The basis for our study was Schwartz's universal human values theory, and we used an associated survey instrument, the Portrait Values Questionnaire (PVQ). Our survey of 193 Bangladeshi female farmers showed that Conformity and Security were regarded as the most important values, while Power, Hedonism and Stimulation were the least important. This finding would be helpful for developers to take into account when developing agriculture apps for this market. In addition, the methodology we used provides a model to follow to elicit the values of apps' users in other communities.*

## 1. Introduction

With the rapid use of smartphones globally, a growing number of organizations have been providing their services and products through mobile software applications (apps). Despite the numerous positive impacts of mobile apps, there is the potential for negative socio-economic consequences, which could stem from violating or ignoring human values such as Privacy, Security, Fairness, Accessibility, Equality, or Tradition [1, 2]. For example, Facebook allowed Cambridge Analytica to use its users' private data for political reasons without users' consents [3] which is a violation of Facebook users' values such as Privacy and Trust. As a result, Facebook had to face humiliation and loss of profit [4]. In another example, Instagram was blamed for the suicide of a British teenager, as Instagram recommended self-harm images to her [5] and violated her values such as Pleasure, Meaning in life and Enjoying life. Values violation in apps can have more destructive impacts on women in conservative societies as they may amplify existing social and cultural challenges [6]. For example, a recent study shows that a huge number of menstruation apps (22 out of 36 apps, 61%) shared users' incredibly sensitive personal details with Facebook without the consents of the users [7, 8]. For the women in conservative societies where these are taboos, we can imagine how much bad impacts this kind of value violations in apps can have on the women's lives and on their mental health.

Several techniques have been proposed to integrate human values in software such as Value-Based Requirement Engineering (VBRE) [9], Value-Sensitive Design (VSD) [10], Value-Sensitive Software Development (VSSD) [11], Values-First SE [12] and Values Q-sort [13]. Despite these efforts, software and apps development companies have limited consideration of the human values of end-users, especially the values of vulnerable groups in conservative societies. It is challenging to consider and integrate human values in software and apps as human values are ambiguous and implicit, and difficult to capture and express [14]. Furthermore, there is also a lack of understanding of what the values of end-users are. We argue that the essential first step to incorporating end-users' values in relevant apps is to understand their values. We have conducted a survey with Bangladeshi female farmers to explore their values. Ensuring that the apps they use are aligned to their values is expected to facilitate the effective and safe participation of women in agriculture and boost the digital economy.

Bangladesh, as a male-dominated conservative society, is heavily dependent on agriculture, which contributes 12.7% of GDP in 2019 [15]. 84% of the country's population earn their livelihood, and 48.1% of the labor force works in the agricultural sector [16]. Rural women in Bangladesh play a significant role in agriculture with rich knowledge and expertise in production [17, 18]. Mobile phones play an important role in developing countries' agriculture by supporting 'sustainable livelihood initiatives' [19]. Providing and equipping rural Bangladeshi women with digital technologies and opportunities such as mobile phones have contributed towards women's empowerment by enabling them to access the information required for successful agricultural initiatives [20]. There were approximately 35 Bangladeshi agriculture applications (apps) available in Google play till September 2018 [21], and female farmers use many of these to provide them with the agricultural information they need [22].

Our survey was completed by 193 female farmers from two Bangladeshi regions. The survey was based on Portrait Values Questionnaire (PVQ) which is a value measurement tool proposed by a social scientist, Shalom H. Schwartz [1]. We found that PVQ has been applied and validated in many countries except Bangladesh [23]. There are also many language version of PVQ except Bengali [24]. Therefore, conducting the case study in Bangladesh would provide an opportunity to compare the results with other countries. The methodology of this research can be replicated to other societies and countries. The main findings of this research were that Conformity and Security were the most important values for Bangladeshi female farmers, and they considered Power, Hedonism, and Stimulation as the least important values. Their value priorities differed according to demographic factors.

## 2. Background and Related Work

### 2.1. Human Values and Values Theory

Human values are people's personal and social preferences reflected in the attitude and behavior of individuals and the functioning of societies [24].

According to Shalom H. Schwartz, "Values are desirable, trans-situational goals, varying in importance, that serve as guiding principles in people's lives" [25]. Social scientists have been conducting research to conceptualize human values since 1950 [25]. For example, Rokeach identified 36 values and they were categorized as goals in life and modes of conduct, which he named terminal values and instrumental values, respectively [26]. Gouveia et al. proposed a three by two framework to categorize values identifying 18 values in six basic value categories [27]. Similarly, through a value measurement analysis, Hofstede proposed four cultural dimensions: power distance, uncertainty avoidance, individualism versus collectivism, and masculinity versus femininity [28]. In 1992, a circular value structure was proposed by Schwartz, which classified 58 human values according to 10 main value categories [29]. The 10 main value categories are defined by their motivational goals as shown in Table 1. We used Schwartz's theory of basic human values in this study. This is the most cited and universally used values theory [21], tested in different settings with variations according to geography, culture, language, religion, politics, age, education and gender [24, 29]. Furthermore, this theory is the most suited for our research as it is widely used in social science research and other disciplines as well [21].

### 2.2. Value Measurement Tools

Schwartz developed the Schwartz Value Survey (SVS) to measure priorities accorded to values, and subsequently developed an additional survey instrument called the Portrait Values Questionnaire (PVQ) [1, 24]. The SVS presents two lists with 30 value items and 26 or 27 value items to the participants to describe potentially desirable end-states and potentially desirable ways of acting respectively. Each value item acts as a guiding principle in life and needs to be rated based on a 9-point scale [1]. The PVQ is based on short text descriptions (portraits) of different people. Each portrait depicts a person's goals, aspirations, and wishes leading to a particular value [1]. Each portrait needs to be rated based on a 6-point scale [24]. PVQ is extensively used to measure human values in different studies. It was used to examine the cross-cultural validity of the Schwartz values theory [24]. PVQ was adopted to introduce European norms and equations for values [30], operationalize personal values in living standards measures [31], identify children's values for educational purposes [32], and explore the relationship between values and personality traits [33].

PVQ is less abstract, more concrete, more context-bound, and less complicated than the SVS, and can be completed in 7 to 10 minutes [24]. Therefore, PVQ is suitable for less-educated people [34] like the participants of our research. In Bangladesh, 68% of farmers' education level is less than Year 10, and 25% have no education [35]. Moreover, the female literacy rate is lower than that of the male in Bangladesh [36]. We, therefore, selected PVQ as our survey instrument to explore the values of Bangladeshi female farmers.

Table 1. Value categories and descriptions [1]

| Value Category | Description (motivational goals) |
|---|---|
| Self-direction | Independent thought and action–choosing, creating, exploring |
| Stimulation | Excitement, novelty, and challenge in life |
| Hedonism | Pleasure or sensuous gratification for oneself |
| Achievement | Personal success through demonstrating competence according to social standards |
| Power | Social status and prestige, control or dominance over people and resources |
| Security | Safety, harmony, and stability of society, of relationships, and of self |
| Conformity | Restraint of actions, inclinations, and impulses likely to upset or harm others and violate social expectations or norms |
| Tradition | Respect, commitment, and acceptance of the customs and ideas that one's culture or religion provides |
| Benevolence | Preserving and enhancing the welfare of those with whom one is in frequent personal contact |
| Universalism | Understanding, appreciation, tolerance, and protection for the welfare of all people and for nature |

### 2.3. Prevalence of Human Values in Apps

Though human values are necessary to be addressed in apps, current software engineering research and practices have limited consideration of human values [14]. Perera et al. conducted a study with 1350 papers from the top software engineering journals and conferences and found that only 16% papers are directly relevant to human values [37]. A recent study also shows that a wide range of human values such as Freedom, Honesty, Security, Privacy and Equality are ignored in the existing Bangladeshi agriculture apps, which causes user dissatisfaction [21].

## 3. Methodology

This research aims to gain a deep understanding of the values of female farmers in Bangladesh. Such knowledge can help (Bangladeshi) agriculture mobile apps' developers consider and integrate the values of this group of end-users when developing agriculture mobile apps. To realize this goal, we formulated the following research questions:

(RQ1) What are the value priorities of female farmers in Bangladesh?

(RQ2) How do Bangladeshi female farmers' value priorities differ demographically?

To answer the research questions, we conducted a survey. Figure 1 provides an overview of the survey as our research method. The details of the survey are described in the following subsections.

### 3.1. Protocol and Measures

Data collection was conducted with a paper-based survey, which took the women approximately 15 minutes to complete. The survey included seven demographics questions (e.g.,*"which age group do you belong to?"*) and 40 questions from the Schwartz Portrait Values Questionnaire (PVQ) [38]. There are several versions of PVQ. In 2001, Schwartz introduced 40 items PVQ (PVQ-40) [38], which has been shortened to 21 items (PVQ-21) in 2002 [39]. Schwartz also developed 20 items and 29 items PVQ [38]. Furthermore, Bubeck and Bilsky developed 29 items PVQ in 2002 [40], and Knoppen and Saris developed 33 items PVQ in 2009 [34]. In this research, we used the 40 items PVQ because it is more suitable for cross-cultural research [39]. It is also argued that this version of the PVQ may lead to more accurate results as it is more elaborate and contains more items in each value [39].

In PVQ, each portrait (i.e., item) describes a person's characteristic that refers to the importance of a particular value. For example, the portrait "*It's very important to her to help the people around her. She wants to care for their wellbeing*" describes a person for whom the value of Benevolence is important. The participants were asked to compare themselves with each portrait and indicate the extent to which they considered the portrait as being like them on a 6-point Likert scale: 6 (Very much like me), 5 (Like me), 4 (Somewhat like me), 3 (A little like me), 2 (Not like me) and 1 (Not like me at all) [24].

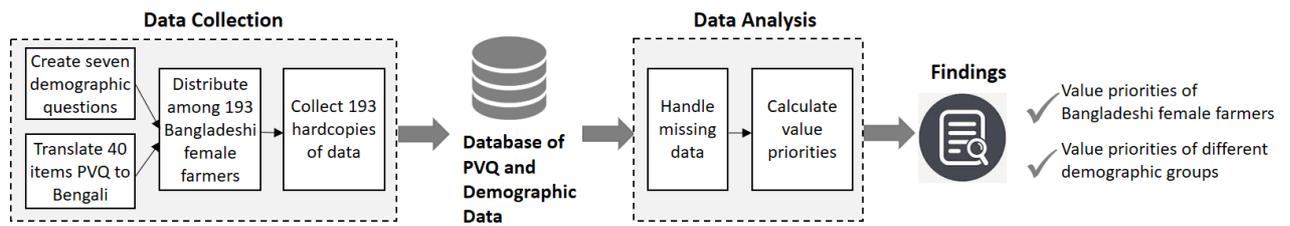

Figure 1. An overview of the research method

As the mother tongue of the participants is Bengali, they should be comfortable to read and write in Bengali. The first author, who is Bangladeshi, translated the survey instrument from English to Bengali. Then we asked an independent translator to translate the Bengali version back to English. It should be noted that the translator did not have any previous knowledge of the English version of PVQ. We then compared the original version of PVQ with the back-translated version and found a few inconsistencies. To address this issue, the first author again translated PVQ to Bengali and the translator was asked to perform the same task. Then, we compared the original version of PVQ with the translated version and found minimal difference between the new version of PVQ-40 and the original. This gave us confidence that the translated Bengali version of PVQ is the same as the original version. Therefore, we conducted the survey with this Bengali version of PVQ.

### 3.2. Participants

A well-recognized multi-national charitable organization, Oxfam, helped us recruit the participants. In one of its projects named PROTIC [41], the organization had provided smartphones to 100 female farmers from the Northern part (sandy area [42]) and 100 from the Southern part (coastal area [43]) of Bangladesh. A local NGO connected with Oxfam trained the 200 female farmers on how to use smartphones and different agriculture mobile applications. The fieldworkers of the NGO introduced the lead author to the farmers. Then, the researcher and fieldworkers explained our research objectives (verbally and in writing) to the participants (200 female farmers). Guidance was provided to the women on how to fill out the survey. It was conducted in their home village environment to help them feel at ease (see Figure 2). Seven women were not available, so we collected data from 193 participants. One of the participants from the Southern part had limited literacy skills. The fieldworkers read the survey questions out to her, she responded orally and they completed the survey on her behalf. Her responses were recorded and checked by the Bangladeshi researcher (first author) to ensure the accuracy of data collection.

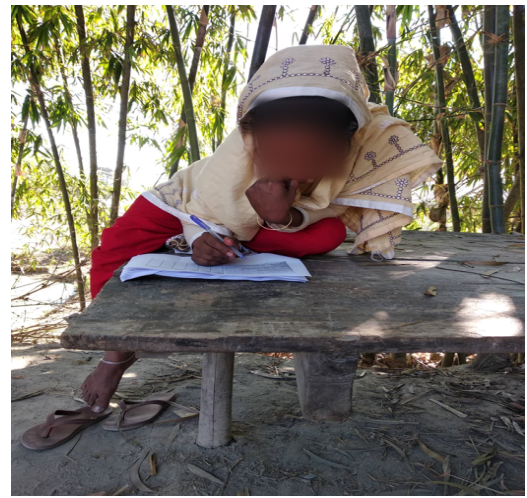

Figure 2. Conducting PVQ in the natural setting

### 3.3. Data Analysis

**Preparation**. Participants' responses recorded on the paper-based survey forms were transferred to SPSS Software [44]. We did not transfer participants' names to ensure that their identities were not disclosed.

**Handling the Missing Data**. We found missing responses for some questions and a few multiple responses for a single question. This could have happened for various reasons. For example, the participant did not understand the question, the question was difficult to answer, and the participant did not feel safe to answer the question [45]. Following the guidelines in [45], we treated the multiple responses as missing responses. We found 275 (3.562%) missing responses out of 7720 (each of 193 participants had to respond to 40 questions in PVQ). We used the *multiple*

Table 2. Different demographic groups and corresponding number of participants

| Total Number of Participants: 193 | | | | | | | | | | | | | | | |
|---|---|---|---|---|---|---|---|---|---|---|---|---|---|---|---|
| Age | | | | | | Education Level | | | | | | | Number of Educated Family Members | | | | |
| <18 | 18-24 | 25-31 | 32-38 | 39-45 | >45 | No schooling | Nursery to 5th grade | 6th to 10th grade | Passed Secondary School Certificate (SSC) | Some college credits but no degree | Passed Higher Secondary Certificate (HSC) | Passed BA/B.Sc. | No one | 1-2 | 3-4 | More than 4 | All of them |
| 0 | 37 | 77 | 53 | 24 | 02 | 0 | 25 | 108 | 17 | 06 | 33 | 04 | 01 | 126 | 43 | 16 | 07 |

*imputation* technique in SPSS to replace the missing data with plausible values [46]. This process enabled us to increase statistical power and avoid getting wrong observations due to reduced sample size [45].

**Calculating Value Priorities**. Each of values in PVQ contains three (e.g., Stimulation) to six (e.g., Universalism) portraits (i.e., items). However, we did not consider the values' average as it might not reflect the importance of the values for the participants [1]. For example, consider two participants who both rated 4 for Benevolence. It may be the most important value for the first participant because she provided ratings lower than 4 for other values. On the other hand, for the second participant, this can be the least important value who rated higher than 4 for other values. To this end, Schwartz [1] proposed measuring value priorities (i.e., the relative importance of the different values). The value priorities were calculated by subtracting each participant's average response to all the value items from her response to each item [1]. To understand the variations of value priorities among different age and education groups, we divided the participants into different age and education groups and calculated their value priorities in the same way. The age and education groups and their corresponding number of participants are shown in Table 2.

## 4. Results

This section presents the results of the analysis of PVQ conducted with 193 Bangladeshi female farmers.

### 4.1. RQ1: Value Priorities of Bangladeshi Female Farmers

Figure 3 shows participants' value priorities. The figure contains all the ten main value categories- Benevolence, Universalism, Self-direction, Stimulation, Hedonism, Achievement, Power, Security, Conformity and Tradition with their priorities. According to Figure 3, Conformity is the most important value for the female farmers, as it has the highest priority value (0.5667). The priority of Security (0.4875) is second ranked. On the other hand, three values are considered less important by the women and have negative priorities: Power (-1.2485), Hedonism (-0.7656) and Stimulation (-0.1121). Among them, Power is valued the least.

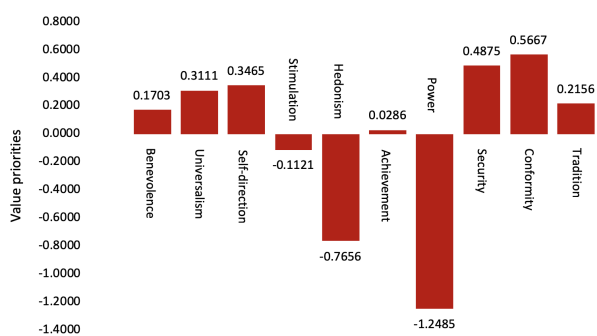

Figure 3. Value priorities of the female farmers in Bangladesh

To better understand the value priorities of the participants, Figure 4 shows the number of positive and negative value priorities for each value. As shown in Figure 4, Conformity has the maximum positive value priorities, as 158 participants (81.87%) out of 193 have positive value priorities, whereas the rest (18.13%) have negative. Similarly, the second-highest positive value priorities occur for Security, which is 140 (72.54%). On the other hand, Power has the lowest positive value priorities. Only 29 participants (15.03%) have positive value priorities, and 164 (84.97%) have negative.

Figure 5 shows how many participants have the same value as their top most priority. We found that Conformity is the top priority for 39 participants, which constitutes the maximum ratio (20.2%). The second highest is Self-direction which is the most important value for 24 participants (12.4%). Security holds the third position (11.4%) whereas in Figure 3, Security is in the second position. This is because, though less number of participants (22) have Security as their most important value than Self-direction, the other participants also gave good ratings to Security. Therefore, the overall value priority becomes higher for Security than Self-direction. 49 participants out of 193 have more than one most important values. For example, Self-direction and Tradition both are the top most priorities for 11 participants that constitutes 5.7%.

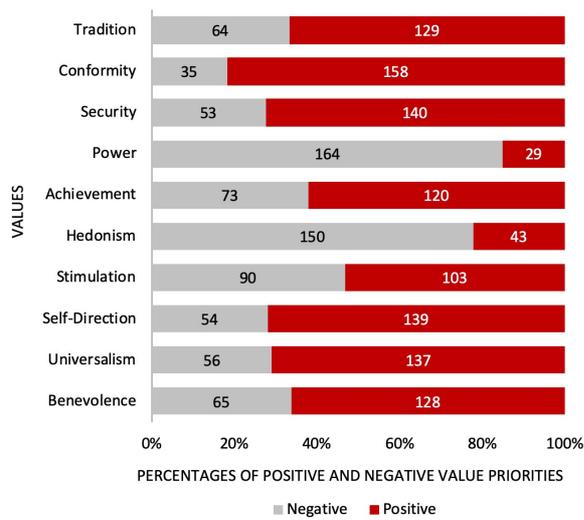

**Figure 4.** Percentages of positive and negative Value priorities

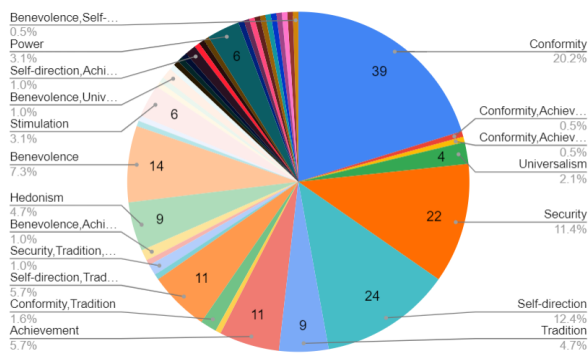

**Figure 5.** Ratios of the most important values of all the participants

### 4.2. RQ2: Differences of Value Priorities in Different Demographics

**Value priorities of different age groups.** The priorities of all the ten value categories for the five different age groups are shown in Table 3. The highest value properties appear in bold, and the lowest value priorities are underlined for each group.

As shown in Table 3, the priorities of Security and Tradition are the highest (0.6579 and 0.2985 respectively) for the lowest age group (18-24 years) and lowest (-0.0575 and -0.7575 respectively) for the highest age group (above 45). In both cases, the priorities become positive to negative as age increases. On the other hand, the priority of Self-direction is the maximum (1.4925) for the highest age group and minimum (0.2714) for the lowest age group. The priority of Self-direction increases with age except for the participants aged 32-38. The priorities of Stimulation increase from negative to positive with the increase of age. The age groups of 39-45 and above 45 have positive priorities of Stimulation (0.1318 and 0.2425 respectively), whereas the other age groups have negative priorities. The priorities of Hedonism and Power are negative for all the age groups. According to Table 3, all the age groups have positive priorities for Benevolence, Universalism, Self-direction, and Conformity, whereas negative for Hedonism and Power.

**Value priorities of different education levels groups.** Table 4 shows the priorities of all the ten value categories for the six groups of different education levels. The highest value priorities are in bold, and the lowest value priorities are underlined for each group.

As shown in Table 4, the priorities of Achievement decreases (positive to negative) as education levels increase. It is the highest (0.3651) for the participants with the lowest level of education (Nursery to 5th grade) and lowest (-0.3271) for the participants who have the highest level of education (Passed BA/BSc). The differences in the priorities of Security with education levels are exactly the opposite of Achievement. It increases with the increase of education levels except for the group of "some college credits but no degree". Security's priority is the lowest (0.0511) for the group with the lowest education level and highest (1.1729) for the group with the highest level of education. Similarly, the priorities of Conformity increase with the increase of education level except for the highest educated group of participants, with a slight decrease from the group who passed their High School Certificate. The priorities of Hedonism and Power are negative for all the groups. The pattern repeats for Stimulation except for the group with some college credits but no degree. Moreover, Table 4 shows that the priorities of Hedonism and Power are negative and the priorities of Universalism, Security, Conformity and Tradition are positive for all six groups.

**Value priorities of the groups of different number of educated family members.** The priorities of the ten value categories for the five groups with different numbers of educated people in the family are shown in Table 5. The highest value properties are bold, and the lowest value priorities are underlined for each group.

The priorities of Hedonism and Power decreases (positive to negative), whereas the priority of Conformity rises with an increase in the number of educated family members. The priorities of Benevolence, Stimulation, and Conformity are the lowest (-0.5133, -0.2633, and 0.2367 respectively) for the least educated family group (No one has education above 6th grade) and highest (0.5009, 0.1914, and 0.8223 respectively) for the most educated family group (all of the family members have education above 6th grade). On the other hand, the priorities of Hedonism and Power are highest (1.07 and 0.07 respectively) for the least educated family group and lowest (-1.0467 and -1.5705 respectively) for the most educated family group. Only Universalism and Conformity have positive value priorities for all the groups.

Table 3. Value priorities of different age groups- the highest value priorities are bold and the lowest value priorities are underlined for each group

|  | Benevolence | Universalism | Self-Direction | Stimulation | Hedonism | Achievement | Power | Security | Conformity | Tradition |
|---|---|---|---|---|---|---|---|---|---|---|
| **18-24** | 0.1633 | 0.4156 | 0.2714 | -0.1205 | -0.6556 | -0.1029 | <u>-1.4898</u> | **0.6579** | 0.5620 | 0.2985 |
| **25-31** | 0.1957 | 0.2953 | 0.3483 | -0.2004 | -0.8887 | 0.1730 | <u>-1.2653</u> | 0.4635 | **0.5951** | 0.2834 |
| **32-38** | 0.2142 | 0.2661 | 0.3180 | -0.1018 | -0.6804 | -0.0971 | <u>-1.2213</u> | 0.5624 | **0.6199** | 0.1199 |
| **39-45** | 0.0068 | 0.2568 | **0.4235** | 0.1318 | -0.6876 | 0.0381 | <u>-0.8960</u> | 0.1818 | 0.3818 | 0.1631 |
| **Above 45** | 0.1175 | 0.8258 | **1.4925** | 0.2425 | <u>-1.2575</u> | 0.1175 | -1.0908 | -0.0575 | 0.3675 | -0.7575 |

Table 4. Value priorities of different groups of education levels- the highest value priorities are bold and the lowest value priorities are underlined for each group

|  | Benevolence | Universalism | Self-Direction | Stimulation | Hedonism | Achievement | Power | Security | Conformity | Tradition |
|---|---|---|---|---|---|---|---|---|---|---|
| **Nursery to 5th grade** | -0.0949 | 0.2417 | 0.2051 | -0.0849 | -0.3783 | **0.3651** | <u>-0.8849</u> | 0.0511 | **0.3651** | 0.2151 |
| **6th to 10th grade** | 0.2216 | 0.3142 | 0.3582 | -0.1441 | -0.7163 | 0.0527 | <u>-1.3015</u> | 0.4755 | **0.5411** | 0.1985 |
| **Passed SSC** | 0.3362 | 0.243 | 0.3803 | -0.2275 | <u>-0.9923</u> | -0.0315 | -0.9726 | **0.6509** | 0.5421 | 0.0715 |
| **Some college credits but no degree** | 0.2967 | 0.2689 | 0.5883 | 0.4633 | -1.1478 | -0.12 | <u>-1.4811</u> | 0.33 | **0.63** | 0.1717 |
| **Passed HSC** | 0.0793 | 0.4278 | 0.4429 | -0.0672 | -1.0268 | -0.204 | <u>-1.4914</u> | 0.7187 | **0.7838** | 0.3369 |
| **Passed BA/BSc** | 0.2979 | 0.0479 | -0.3896 | -0.1604 | -0.8271 | -0.3271 | <u>-0.9104</u> | **1.1729** | 0.7354 | 0.3604 |

## 5. Discussion

### 5.1. Analysis of the Results

We found that Conformity was the most important value for Bangladeshi female farmers. This reflects the social environment of Bangladesh. The participants belong to a patriarchal society [6], where living in husband's extended family, honoring elders, and being polite and obedient [47] are essential cultural characteristics. The second most important value for the farmers was Security, reflecting their vulnerable status given the prevalence of violence against women in Bangladesh [48]. The women accorded the least importance to Power, which again reflects social norms. In Bangladesh, marginalized women are dependent on the male figures or the senior members of their family even for small things [49] and are therefore unlikely to expect to exercise any power. Hedonism was also considered of lesser importance. A research shows that families expect new brides to spend their time learning their new domestic responsibilities, rather than indulging in talking or socialising with neighbours [50].

We observed that Hedonism correlates negatively with age, which is in line with the findings of [24, 29, 51], supposedly because older people are less interested in enjoying life, pleasure, and challenges [52, 53]. Similarly, our findings regarding the positive correlation between Self-direction and level of education (the participants with the highest education level were an exception) confirm the results of [24, 29, 51]. One reason for this could be that education helps people be more open, flexible, creative, and concerned about self-direction values [24]. However, in Bangladesh, young women are more vulnerable to violence than older women [54]. Our research results also show that the priority of Security is the highest for the youngest participants and lowest for the oldest participants. We observed the same pattern for Tradition. Young Bangladeshi women are more responsible for maintaining social norms and traditions, such as looking after the husband's extended family [55]. The negative correlations of the priorities of Security and Tradition with age in our study is not in line with the findings of [24, 29, 51]. All this indicates that values can be different for different groups of people, even within the same societies [29]. We argue that there is a need for further research to investigate the values of individuals in different domains and societies, particularly more vulnerable or silent people.

The granularity of values in the Schwartz model can be used to better inform app developers. For example, there are four categories under Conformity, namely Honoring of elders, Politeness, Obedient and Self-discipline [1]. As Conformity is the most important value of the Bangladeshi female farmers, app developers may use each of these items to further improve the design choices when they develop apps. However, a study which analysed users' reviews of Bangladeshi agriculture apps found that they are unable to deliver all the desired values of users [21]. For example, though Security and Conformity are the most important values of our target people, these are not carefully considered in the existing Bangladeshi apps. Al-Ameen et al. explained how vulnerable Bangladeshi apps are and the importance of Security for Bangladeshi apps users [8]. There are several value items under Conformity such as politeness. Lack of politeness in apps can be a reason for users' dissatisfaction. Carolus et al. described that the participants in their study provided negative reviews for the apps speaking impolitely [56]. As Bangladeshi female farmers are increasingly using agriculture apps to participate in agriculture and contribute to the economy [22], their values should be reflected in the apps. Researchers, practitioners, and policymakers can leverage our findings to provide a safe and comfortable environment for female farmers to participate effectively in agriculture and contribute to the digital economy.

Table 5. Value priorities of the groups with different numbers of educated family members- the highest value priorities are bold and the lowest value priorities are underlined for each group

|  | Benevolence | Universalism | Self-Direction | Stimulation | Hedonism | Achievement | Power | Security | Conformity | Tradition |
|---|---|---|---|---|---|---|---|---|---|---|
| **No one** | <u>-0.5133</u> | 0.4033 | <u>-0.5133</u> | -0.2633 | **1.07** | -0.0133 | 0.07 | -0.4633 | 0.2367 | -0.0133 |
| **1 to 2** | 0.159 | 0.2965 | 0.2205 | -0.1453 | -0.6834 | 0.153 | <u>-1.2141</u> | 0.4336 | **0.4943** | 0.2859 |
| **3 to 4** | 0.1694 | 0.2721 | 0.5764 | -0.1697 | -0.9062 | 0.0357 | <u>-1.2395</u> | **0.6845** | 0.652 | -0.0748 |
| **More than 4** | 0.1601 | 0.3997 | 0.6757 | 0.1809 | -1.0274 | -0.6899 | <u>-1.4857</u> | 0.5601 | **0.8164** | 0.4101 |
| **All of them** | 0.5009 | 0.5961 | 0.5723 | 0.1914 | -1.0467 | -0.6062 | <u>-1.5705</u> | 0.218 | **0.8223** | 0.3223 |

### 5.2. Implications for App Development

We provide a prioritized list of value categories as a basis for app developers to align the Bangladeshi agriculture apps with the values of female farmers. However, these values should not be considered as an afterthought but should be incorporated early during the app design, when the system is still malleable [57]. In this study, PVQ was successful as a values elicitation technique. We propose to use our approach to elicit values from users to develop value requirements for the app development life-cycle. These values can be fed as an input to the readily available techniques dedicated to incorporate users in the requirement, design and development process of a system such as Value-Based Requirement Engineering (VBRE) [9], Participatory Design (PD) [58], Value-Sensitive Design (VSD) [10], Value-Sensitive Software Development (VSSD) [11], Values-First SE [12] and Values Q-Sort [13].

Our results reveal that the importance of values varies in different demographics. For example, Self-direction is the most important value for the older age group of participants, while the youngest group perceives Conformity and Security as the most important values. Given the diverse demographics that use agriculture apps and possible conflicts between values, the apps should be designed to be adaptable to support different (and sometimes conflicting) values from different demographics. More research should focus on the role of app developers to cater the diversity of users.

### 6. Conclusions, Limitations and Future Work

**Conclusions.** This study collected empirical data to explore the values of Bangladeshi female farmers in different age groups and with differing levels of education through a survey completed by 193 Bangladeshi female farmers. We found that Conformity and Security were the most important values, whereas Power, Hedonism and Stimulation were the least important for Bangladeshi female farmers. Their values differ slightly with different demographics. Conformity and Security are their most important values except for the older age, lowest education level and least educated family groups. On the other hand, Power and Hedonism are the least important values for all age, education level and educated family groups except for the least educated family group. Our study's findings are expected to help Bangladeshi software development organizations and practitioners develop agriculture apps that address the values of the Bangladeshi female farmers.

**Limitations.** The main limitation of this study is that the results cannot be generalized to all female farmers. However, we designed the research method in such a way to allow the replication and validation of this study in different contexts. There is a risk that the version of PVQ-40 we used is not the same as the original version as we needed to translate it to Bengali. However, we minimized this risk by translating and back-translating twice. There is also another risk of not getting the accurate results for having lower sample size for highest age group (2 participants) and least educated family group (1 participant).

**Future Work.** As we conducted our survey in both the Northern and Southern parts of Bangladesh, we plan to explore how cultural differences impact value priorities. We also aim to conduct a similar study with

Bangladeshi male farmers to determine the extent to which the value priorities are gender-specific.

## References

[1] S. H. Schwartz, "An overview of the schwartz theory of basic values," *Online Readings in Psychology and Culture*, vol. 2, no. 1, pp. 2307–0919, 2012.

[2] J. Whittle, M. A. Ferrario, W. Simm, and W. Hussain, "A case for human values in software engineering," *IEEE Software*, 2019.

[3] C. Cadwalladr and E. Graham-Harrison, "Revealed: 50 million facebook profiles harvested for cambridge analytica in major data breach." https://www.theguardian.com/news/2018/mar/17/cambridge-analytica-facebook-influence-us-election, March 2018. Accessed: 2019-10-28.

[4] R. Neate, "Over $119bn wiped off facebook's market cap after growth shock." https://www.theguardian.com/technology/2018/jul/26/facebook-market-cap-falls-109bn-dollars-after-growth-shock, July 2018. Accessed: 2019-10-28.

[5] A. Crawford, "Instagram 'helped kill my daughter'." https://www.bbc.com/news/av/uk-46966009/instagram-helped-kill-my-daughter, January 2019. Accessed: 2019-10-28.

[6] S. Sultana, F. Guimbretière, P. Sengers, and N. Dell, "Design within a patriarchal society: Opportunities and challenges in designing for rural women in bangladesh," in *Proceedings of the 2018 CHI Conference on Human Factors in Computing Systems*, pp. 1–13, 2018.

[7] "No body's business but mine: How menstruation apps are sharing your data." https://www.privacyinternational.org/long-read/3196/no-bodys-business-mine-how-menstruations-apps-are-sharing-your-data, September 2019. Accessed: 2019-09-09.

[8] M. N. Al-Ameen, T. Tamanna, S. Nandy, M. M. Ahsan, P. Chandra, and S. I. Ahmed, "We don't give a second thought before providing our information: Understanding users' perceptions of information collection by apps in urban bangladesh," in *Proceedings of the 3rd ACM SIGCAS Conference on Computing and Sustainable Societies*, pp. 32–43, 2020.

[9] S. Thew and A. Sutcliffe, "Value-based requirements engineering: method and experience," *Requirements engineering*, vol. 23, no. 4, pp. 443–464, 2018.

[10] J. Davis and L. P. Nathan, "Value sensitive design: Applications, adaptations, and critiques," *Handbook of ethics, values, and technological design: Sources, theory, values and application domains*, pp. 11–40, 2015.

[11] H. Aldewereld, V. Dignum, and Y.-h. Tan, "Design for values in software development," *Handbook of Ethics, Values, and Technological Design: Sources, Theory, Values and Application Domains*, pp. 831–845, 2015.

[12] M. A. Ferrario, W. Simm, S. Forshaw, A. Gradinar, M. T. Smith, and I. Smith, "Values-first se: research principles in practice," in *2016 IEEE/ACM 38th International Conference on Software Engineering Companion (ICSE-C)*, pp. 553–562, IEEE, 2016.

[13] E. Winter, S. Forshaw, and M. A. Ferrario, "Measuring human values in software engineering," in *Proceedings of the 12th ACM/IEEE International Symposium on Empirical Software Engineering and Measurement*, pp. 1–4, 2018.

[14] D. Mougouei, H. Perera, W. Hussain, R. Shams, and J. Whittle, "Operationalizing human values in software: A research roadmap," in *Proceedings of The 2018 26th ACM Joint Meeting on European Software Engineering Conference and Symposium on The Foundations of Software Engineering*, pp. 780–784, ACM, 2018.

[15] T. W. Bank, "Agriculture, forestry, and fishing, value added (% of gdp)." https://data.worldbank.org/indicator/NV.AGR.TOTL.ZS, 2019. Accessed: 2020-02-26.

[16] M. M. M. Uddin, "Causal relationship between agriculture, industry and services sector for gdp growth in bangladesh: An econometric investigation," *Journal of Poverty, Investment and Development*, vol. 8, 2015.

[17] S. Rahman, "Women's labour contribution to productivity and efficiency in agriculture: empirical evidence from bangladesh," *Journal of Agricultural Economics*, vol. 61, no. 2, pp. 318–342, 2010.

[18] R. Aktar, A. Chowdhury, A. Zakaria, C. Vogl, et al., "Seed information and communication networks of male and female farmers: A micro level study in bangladesh," *Building Sustainable Rural Futures. The Added Value of Systems Approaches in Times of Change and Uncertainty*, pp. 760–769, 2010.

[19] M. K. Islam and F. Slack, "Women in rural bangladesh: Empowered by access to mobile phones," in *Proceedings of the 9th International Conference on Theory and Practice of Electronic Governance*, pp. 75–84, 2016.

[20] L. Stillman, M. Sarrica, and T. Denison, "After the smartphone has arrived in the village. how practices and proto-practices emerged in an ict4d project.," *11th International Development Informatics Association Conference*, 2020.

[21] R. A. Shams, W. Hussain, G. Oliver, A. Nurwidyantoro, H. Perera, and J. Whittle, "Society-oriented applications development: Investigating users' values from bangladeshi agriculture mobile applications," in *Proceedings of The 42nd International Conference on Software Engineering*, ACM, 2020.

[22] M. S. Rahman, M. E. Haque, and M. S. I. Afrad, "Utility of mobile phone usage in agricultural information dissemination in bangladesh," *East African Scholars Journal of Agriculture and Life Sciences*, vol. 3, 2020.

[23] L. V. Romanyuk and Y. Chernov, "Personality values and motivations in cross-cultural context: Assessment by pvq-test and handwriting psychology," *Science and Education*, 2017.

[24] S. H. Schwartz, G. Melech, A. Lehmann, S. Burgess, M. Harris, and V. Owens, "Extending the cross-cultural validity of the theory of basic human values with a different method of measurement," *Journal of Cross-cultural Psychology*, vol. 32, no. 5, pp. 519–542, 2001.

[25] S. H. Schwartz, "Basic human values: An overview," *Recuperado de http://www. yourmorals. org/schwartz*, 2006.

[26] M. Rokeach, "The nature of human values.," pp. 359–361, 1973.

[27] V. V. Gouveia, T. L. Milfont, and V. M. Guerra, "Functional theory of human values: Testing its content and structure hypotheses," *Personality and Individual Differences*, vol. 60, pp. 41–47, 2014.

[28] G. Hofstede and M. H. Bond, "Hofstede's culture dimensions: An independent validation using rokeach's value survey," *Journal of Cross-cultural Psychology*, vol. 15, no. 4, pp. 417–433, 1984.

[29] S. H. Schwartz, "Universals in the content and structure of values: Theoretical advances and empirical tests in 20 countries," *Advances in Experimental Social Psychology*, vol. 25, no. 1, pp. 1–65, 1992.

[30] M. Verkasalo, J.-E. Lönnqvist, J. Lipsanen, and K. Helkama, "European norms and equations for a two dimensional presentation of values as measured with schwartz's 21-item portrait values questionnaire," *European Journal of Social Psychology*, vol. 39, no. 5, pp. 780–792, 2009.

[31] L. M. Ungerer and J. Joubert, "The use of personal values in living standards measures," *Southern African Business Review*, vol. 15, no. 2, 2011.

[32] A. K. Döring, "Assessing children's values: An exploratory study," *Journal of Psychoeducational Assessment*, vol. 28, no. 6, pp. 564–577, 2010.

[33] J. Saiz, J. L. Alvaro, and I. Martinez, "Relation between personality traits and personal values in cocaine-dependent patients," *Adicciones*, vol. 23, no. 2, pp. 125–132, 2011.

[34] C. Beierlein, E. Davidov, P. Schmidt, S. H. Schwartz, and B. Rammstedt, "Testing the discriminant validity of schwartz'portrait value questionnaire items–a replication and extension of knoppen and saris (2009)," in *Survey Research Methods*, vol. 6, pp. 25–36, 2012.

[35] M. S. Islam and Å. Grönlund, "Factors influencing the adoption of mobile phones among the farmers in bangladesh: Theories and practices," *International Journal on Advances in ICT for Emerging Regions*, vol. 4, no. 1, 2011.

[36] J. Ferdaush and K. Rahman, "Gender inequality in bangladesh," *Report of Enhancing The Responsiveness of The Government to Address Exclusion and Inequality*, 2011.

[37] H. Perera, W. Hussain, J. Whittle, A. Nurwidyantoro, D. Mougouei, R. A. Shams, and G. Oliver, "A study on the prevalence of human values in software engineering publications, 2015 – 2018," in *Proceedings of The 42nd International Conference on Software Engineering*, ACM, 2020.

[38] S. H. Schwartz, "A proposal for measuring value orientations across nations," *Questionnaire Package of The European Social Survey*, vol. 259, no. 290, p. 261, 2003.

[39] J. Cieciuch and E. Davidov, "A comparison of the invariance properties of the pvq-40 and the pvq-21 to measure human values across german and polish samples," in *Survey Research Methods*, vol. 6, pp. 37–48, 2012.

[40] M. Bubeck and W. Bilsky, "Value structure at an early age," *Swiss Journal of Psychology*, vol. 63, no. 1, pp. 31–41, 2004.

[41] M. Sarrica, T. Denison, L. Stillman, T. Chakraborty, and P. Auvi, ""what do others think?" an emic approach to participatory action research in bangladesh," *AI & SOCIETY*, vol. 34, no. 3, pp. 495–508, 2019.

[42] S. I. Anik and M. A. S. A. Khan, "Climate change adaptation through local knowledge in the north eastern region of bangladesh," *Mitigation and Adaptation Strategies for Global Change*, vol. 17, no. 8, pp. 879–896, 2012.

[43] M. Rakib, J. Sasaki, S. Pal, M. A. Newaz, M. Bodrud-Doza, and M. A. Bhuiyan, "An investigation of coastal vulnerability and internal consistency of local perceptions under climate change risk in the southwest part of bangladesh," *Journal of Environmental Management*, vol. 231, pp. 419–428, 2019.

[44] S. Landau, "A handbook of statistical analyses using spss," pp. 12–33, 2004.

[45] M. Vriens and S. Sinharay, "Dealing with missing data in surveys and databases," *The Handbook of Marketing Research: Uses, Misuses, and Future Advances*, p. 178, 2006.

[46] H. Zhong, W. Hu, and J. M. Penn, "Application of multiple imputation in dealing with missing data in agricultural surveys: The case of bmp adoption," *Journal of Agricultural and Resource Economics*, vol. 43, no. 1835-2018-707, pp. 78–102, 2018.

[47] T. M. Islam, M. I. Tareque, A. D. Tiedt, and N. Hoque, "The intergenerational transmission of intimate partner violence in bangladesh," *Global Health Action*, vol. 7, no. 1, p. 23591, 2014.

[48] I. Dankelman, "Gender, climate change and human security lessons from bangladesh, ghana and senegal," *The Women's Environment and Development Organization*, 2008.

[49] S. R. Rashid, "Bangladeshi women's experiences of their men's migration: Rethinking power, agency, and subordination," *Asian Survey*, vol. 53, no. 5, pp. 883–908, 2013.

[50] S. Amin, L. Suran, *et al.*, "Terms of marriage and time-use patterns of young wives–evidence from rural bangladesh," *Electronic International Journal of Time Use Research*, vol. 6, no. 1, pp. 92–108, 2009.

[51] G.-R. M. Rosario, D.-F. M. Carmen, and S. Biagio, "Values and corporate social initiative: An approach through schwartz theory," *International Journal of Business and Society*, vol. 15, no. 1, p. 19, 2014.

[52] N. D. Glenn, "Aging and conservatism," *The ANNALS of the American Academy of Political and Social Science*, vol. 415, no. 1, pp. 176–186, 1974.

[53] T. R. Tyler and R. A. Schuller, "Aging and attitude change.," *Journal of Personality and Social Psychology*, vol. 61, no. 5, p. 689, 1991.

[54] H. Zaman, "Violence against women in bangladesh: issues and responses," in *Women's Studies International Forum*, vol. 22, pp. 37–48, Elsevier, 1999.

[55] D. Balk, "Defying gender norms in rural bangladesh: A social demographic analysis," *Population Studies*, vol. 51, no. 2, pp. 153–172, 1997.

[56] A. Carolus, R. Muench, C. Schmidt, and F. Schneider, "Impertinent mobiles-effects of politeness and impoliteness in human-smartphone interaction," *Computers in Human Behavior*, vol. 93, pp. 290–300, 2019.

[57] C. Detweiler and M. Harbers, "Value stories: Putting human values into requirements engineering.," in *REFSQ Workshops*, pp. 2–11, 2014.

[58] M. J. Muller and S. Kuhn, "Participatory design," *Communications of the ACM*, vol. 36, no. 6, pp. 24–28, 1993.